\documentclass[apj]{emulateapj}

\shorttitle{Mass-Radius Constraints for X7 and X5 in 47 Tuc}
\shortauthors{Bogdanov et al.}

\submitted{Submitted to the Astrophysical Journal on March 4, 2016}

\begin{document}

\title{Neutron Star Mass-Radius Constraints of the Quiescent Low-mass \\ 
X-ray Binaries  X7 and X5 in the Globular Cluster 47 Tuc}

\author{Slavko Bogdanov\altaffilmark{1}, Craig
  O.~Heinke\altaffilmark{2}, Feryal \"Ozel\altaffilmark{3}, Tolga
  G\"uver\altaffilmark{4}}

\altaffiltext{1}{Columbia Astrophysics Laboratory, Columbia
  University, 550 West 120th Street, New York, NY, 10027, USA}

\altaffiltext{2}{Department of Physics, University of Alberta, CCIS
  4-183, Edmonton, AB T6G 2E1, Canada}

\altaffiltext{3}{Department of Astronomy, University of Arizona, 933
  North Cherry Avenue, Tucson, AZ, 85721, USA}

\altaffiltext{4}{Istanbul University, Science Faculty, Department of
  Astronomy and Space Sciences, Beyazit, 34119, Istanbul, Turkey}

\begin{abstract}
We present \textit{Chandra} ACIS-S sub-array observations of the
quiescent neutron star low-mass X-ray binaries X7 and X5 in the
globular cluster 47~Tuc. The large reduction in photon pile-up
compared to previous deep exposures enables a substantial improvement
in the spectroscopic determination of the neutron star radius and mass
of these neutron stars. Modeling the thermal emission from the neutron
star surface with a non-magnetized hydrogen atmosphere and accounting
for numerous sources of uncertainties, we obtain for the neutron star
in X7 a radius of $R=11.1^{+0.8}_{-0.7}$~km for an assumed stellar
mass of $M=1.4$ M$_{\odot}$ (68\% C.L.). We argue, based on
astrophysical grounds, that the presence of a He atmosphere is
unlikely for this source. Due to eclipses and variable absorption, the
quiescent low-mass X-ray binary X5 provides less stringent
constraints, leading to a radius of $R=9.6^{+0.9}_{-1.1}$~km, assuming
a hydrogen atmosphere and a mass of $M=1.4$ M$_{\odot}$. When combined
with all other existing spectroscopic radius measurements, these
measurements strongly favor radii in the $9.9-11.2$~km range for a
$\sim$1.5 M$_{\odot}$ neutron star and point to a dense matter
equation of state that is somewhat softer than the nucleonic ones that
are consistent with laboratory experiments at low densities.
\end{abstract}

\keywords{dense matter --- equation of state --- globular clusters:
  individual (47 Tucanae) --- stars: neutron}

\section{Introduction}
The equation of state of cold, stable matter at densities that exceed
the nuclear saturation density ($\rho_{\rm sat} = 2.8 \times
10^{14}$~g~cm$^{-3}$) remains one of the principal outstanding
problems in nuclear physics. Neutron stars (NSs) provide a unique
setting where the properties of neutron-rich matter at extreme
conditions can be probed. This is because the mass-radius ($M-R$)
relation of NSs is determined by the dense matter equation of state
(EoS) and, in turn, measuring the mass $M$ and radius $R$ of several
NSs with $<$10\% errors can place strong limits on the EoS at high
densities \citep[see, e.g.,][]{Oz10,Stein10}.

It is well established that observing thermal radiation from the
surface of a NS can serve as a useful tool towards this end.  For a NS
radiating uniformly from its entire surface, one can derive
constraints on its mass and radius by fitting its spectrum with an
appropriate atmosphere model -- when the surface composition is known
or can be determined from the X-ray spectrum itself -- and combining
the spectroscopic measurements with the distance to the source. Low
magnetic field ($\ll$$10^{10}$ G) sources are typically chosen for
such studies so that the radiation transport or temperature
distribution on the stellar surface are not affected by the magnetic
field \citep[see reviews by ][]{Oz13,Pot14}. These criteria are met in
quiescent low-mass X-ray binaries (qLMXBs) containing NSs and, in
particular, for those located in globular clusters, to which the
distances are well-constrained \citep{Rut02}.  In these systems, the
heat stored in the NSs is believed to be deposited by nuclear fusion
in the deep crust during accretion, and is reradiated from the whole
surface when accretion ceases, producing a long-lived thermal glow
\citep{Brown98,Camp98}.  Their low magnetic fields ($\sim$$10^{8-9}$
G) as well as their primarily thermal spectra make qLMXBs fairly clean
laboratories for studies of fundamental NS physics. As such, they can
provide potentially strong constraints on NS structure that are
complementary to those obtained using other techniques.

The observed thermal X-ray radiation from qLMXBs is modeled by
light-element atmospheres because the lightest element that is present
floats to the top of the atmosphere due to rapid gravitational
settling on neutron star surfaces. Light-element atmospheres shift the
peak of the emitted radiation to higher energies compared to a Planck
spectrum of the same effective temperature because of the strong
dependence of free-free absorption on photon energy and the large
temperature gradients in such atmospheres \citep{Raj96,Zavlin96}. By
applying the non-magnetic NS hydrogen atmosphere models of
\citet{Zavlin96} to X-ray observations of qLMXBs,
\citet{Rut99,Rut01a,Rut01b} obtained the first broad constraints on
their radii. These earlier analyses of X-ray spectra from neutron
stars with hydrogen atmosphere models produced radius estimates that
are in reasonable agreement with theoretical predictions. These
findings motivated a host of subsequent observations primarily using
\textit{Chandra} and, to a lesser extent, \textit{XMM-Newton}, in an
attempt to place tighter constraints on the NS radius. The known
qLMXBs in globular clusters, in particular, have been subject of
intensive studies, including X7 and X5 in 47 Tuc
\citep{Heinke03,Heinke06}, U24 in NGC 6397 \citep{Gui11,Heinke14},
source 26 in M28 \citep{Beck03,Serv12}, NGC 2808
\citep{Webb07,Serv08}, M13 \citep{Gend03a,Webb07,Cat13}, $\omega$
Centauri \citep{Rut02,Gend03b,Heinke14}, and M30 \citep{Lug07,Gui14}.

Radius constraints obtained for qLMXBs from the different analyses
have been combined and used for constraining the neutron star
EoS. \citet{Gui13} applied Bayesian analysis techniques on the data
for five qLMXBs to determine the combined mass-radius relation. In
this study, a number of sources of measurement uncertainty were
incorporated into the final constraints, but some systematic
uncertainties were not explored \citep{Lat14,Heinke14}.  These include
the possibility of a helium instead of hydrogen atmosphere for a
subset of the sources as well as the effects of the model used for the
relative abundances of the intervening absorbing material on the
derived radius constraints.  In a subsequent study of six qLMXBs,
including those from \citet{Gui13}, \citet{Oz15} carried out a
uniform analysis of these sources, taking into account the additional
sources of uncertainty and found good agreement between all
measurements. Furthermore, combining the qLMXB
measurements with the radius measurements obtained during
thermonuclear bursts led to a combined NS radius of $10.1-11.1$~km
(95\% C.L.) and to tight constraints on the dense matter equation of
state. This comprehensive study, however, excluded two of the earlier
qLMXB radius constraints, namely those of X5 and X7
\citep{Heinke03,Heinke06} due to concerns in the quality of the
spectral data in these earlier {\it Chandra} observations, as we
describe below.

The globular cluster 47 Tuc (NGC 104) hosts X7 and X5, with the
highest and second-highest X-ray flux at Earth of any qLMXB in a
globular cluster, making them well-suited targets for NS EoS
constraints. A 270-ks \textit{Chandra} observation of 47 Tuc in 2002
\citep{Heinke05} produced a spectrum of X7 with a high number of
counts, which allowed the first measurements on the mass and radius of
this source \citep{Heinke06}.  Using the {\tt nsatmos} hydrogen
atmosphere model in {\tt XSPEC}, for an assumed NS mass of 1.4
M$_{\odot}$, the stellar radius was constrained to be
$14.5^{+1.8}_{-1.6}$ km (90\% C.L.).  However, because the 270-ks
\textit{Chandra} ACIS-S exposure of 47~Tuc was obtained in full-frame
mode with a read-out time of 3.2 seconds, X7 suffered from strong
event pile-up.\footnote{Pile-up occurs when two or more photons,
  arriving at the same detector pixel during one frame time, are
  erroneously identified as a single photon with the sum of the photon
  energies or else altogether discarded \citep{Davis01}. The result is
  a distortion of the intrinsic shape of the source spectrum.} This
resulted in substantial degradation in the quality of the
spectrum. Furthermore, an extra model component to account for pile-up
was required when fitting the spectrum, with a pile-up parameter that
was estimated to be 15\% in the 2002 full-frame dataset.

It is difficult to quantify the systematic uncertainty introduced into
the radius measurement by pile-up for two reasons. First, even though
the pile-up correction is theoretically sound and verified to give
reasonable results, is not as well-calibrated as \textit{Chandra}'s
performance in the absence of pile-up, especially when the pile-up
fraction is as high as 15\% \citep{Stein10}. Second, the inferred
radius is highly sensitive to the spectral shape, which is not
possible to fully correct for. In light of this, it is clear that
minimizing pile-up is critical for obtaining reliable constraints on
the NS $M-R$ relation with qLMXBs.

X5, an X-ray binary that is viewed edge-on \citep{Heinke03}, on the
other hand, shows eclipses with an 8.7-hour period and irregular
energy-dependent dipping. The hydrogen column density $N_{\rm H}$ to
X5 was high and variable during the 2002 ACIS-S observations but less
so during the 2000 ACIS-I and 2005 HRC-S observations.  The accretion
disk in X5 appears to precess, occasionally blocking the view of the
NS, as it did in 2002. Therefore, a clear view of the NS was necessary
to obtain an improved $M-R$ measurement for this target.

\begin{deluxetable}{crccc}
\tablewidth{0pt}
\tablecaption{\textit{Chandra} ACIS Sub-array Observations of 47 Tuc used in this study.}
\tablehead{
\colhead{Instrument} & \colhead{ObsID} & \colhead{Date} & \colhead{Array} & \colhead{Exposure}  \\
 \colhead{}      & \colhead{}      & \colhead{(UT)} & \colhead{size} & \colhead{(ks)}}
\startdata
ACIS-I &  78  & 2000 Mar 16 & 1/4  &  3.9  \\
ACIS-S &  3384  & 2002 Sep 30 & 1/4  & 5.3  \\
ACIS-S &  3385  & 2002 Oct 01 & 1/4  & 5.3  \\
ACIS-S &  3386  & 2002 Oct 03 & 1/4  & 5.5  \\
ACIS-S &  3387  & 2002 Oct 11 & 1/4  & 5.7  \\
ACIS-S &  15747  & 2014 Sep 09 & 1/8  & 50.0  \\
ACIS-S &  15748  & 2014 Oct 02 & 1/8  & 16.2  \\
ACIS-S &  16527  & 2014 Sep 05 & 1/8  & 40.9  \\
ACIS-S &  16528  & 2015 Feb 02 & 1/8  & 40.3  \\
ACIS-S &  16529  & 2014 Sep 21 & 1/8  & 24.7  \\
ACIS-S &  17420  & 2014 Sep 30 & 1/8  & 9.1 
\enddata
\label{logtable}							 
\end{deluxetable}

Here, we present new \textit{Chandra} ACIS-S subarray observations of
X5 and X7, totaling 181~ks, optimized for NS EoS constraints via X-ray
spectroscopy. We show that the data suffer minimally from pile-up in
the new observations and lead to radius constraints that are
significantly different from those of the earlier studies of X5 and
X7. Furthermore, we show that the new radius measurements are in
excellent agreement with those from other qLMXBs and thermonuclear
bursters.

The paper is organized as follows. In \S2 we describe the observations
and data reduction procedure. In \S3, we assess the relevance of a
number of causes of measurement bias and error, while we present
the results of our spectroscopic analysis in \S4. In \S5 we discuss
the implications of our findings and offer conclusions in \S6.

\section{Observations}
For the analysis presented here, we focus on new \textit{Chandra}
ACIS-S exposures totaling 181~ks that were acquired between 2014
September and 2015 February. The observations are summarized in
Table~1. During the exposures, only the ACIS-S3 chip was active and
was configured in a custom 1/8 sub-array, which restricts the region
of the CCD in which data is taken to 128 rows, starting at row 449
(inclusive). This 1/8 subarray affords a 0.4~s frame time, which
dramatically reduces the pile-up fraction of moderately bright sources
such as X7 and X5.

The data extraction was carried out using CIAO\footnote{Chandra
  Interactive Analysis of Observations.} \citep{Fruscione06} version
4.7 and the corresponding calibration database (CALDB 4.6.7).  For
each observation, the source counts were extracted from circular
regions of radius 2.5$''$ centered on the {\tt wavdetect}-derived
positions of both X5 and X7, which enclose $\approx$95\% of the total
source energy. The background was obtained from a source-free region
on the image. The spectra from the individual observations were
generated using {\tt specextract} and were co-added using the {\tt
  combine\_spectra} task in CIAO.  Due to known spectral calibration
issues with ACIS-S data around 0.4 keV, we restrict our
spectroscopic analysis to energies $\ge$0.5 keV.

Using the CIAO tool {\tt pileup\_map}, we determined that the highest
count rates per frame were 0.029 and 0.025 for X7 and X5, respectively,
corresponding to a pile-up level of $\approx$1\%.  While there are
archival sub-array observations from 2000 and 2002, they were acquired
in a 1/4 array size, which provides a 0.8 s readout time and hence a
higher pile-up fraction. Based on this and the short total exposure of
these data (26 ks), we do not consider them in the spectral analysis
used to derive the $M-R$ constraints presented below.  We do, however,
make use of these observations to examine the long-term variability of
X7.

\section{Sources of Measurement Bias}
In order to utilize qLMXBs as high-precision probes of NS structure,
it is important to explore and quantify the effect of any instrument
or modeling uncertainties on the desired NS $M-R$ measurement.
Following the investigations presented in \citet{Heinke14} and
\citet{Oz15}, we examine in this section several potentially important
sources of measurement uncertainty and bias that may affect the EOS
constraints obtained for X7 and X5.

\subsection{Variability}
Rapid (seconds/hours) and long-term ($\sim$years) variability in
qLMXBs is often taken as an indicator of on-going low-level accretion
onto the neutron star. In such an event, the assumptions of a
steady-state, passively cooling atmosphere, a purely thermal flux from
the surface, or a uniformly hot star may no longer be valid.
\citet{Walsh15} and \citet{Bah15} have evaluated the variability of
the thermal emission from 9 and 12 qLMXBs, respectively \citep[see
  also][for individual sources]{Heinke06,Gui11,Heinke14}. For the 7
qLMXBs with purely thermal spectra, there is no evidence for
temperature variations over $\sim$10 years down to levels of 11\%.

The multi-epoch and high-quality \textit{Chandra} ACIS data set of X7
spanning $\sim$15 years allows us to place substantially stricter
limits on the variability of its thermal radiation using spectral fits
to the 2014/2015 combined data, and to the 2000 and 2002 subarray
data\footnote{Because the extensive ($\sim$800 ks) \textit{Chandra}
  HRC-S data set provides no spectral information, it is not useful in
  this regard.}.  For the most stringent test, we kept all parameters
fixed except for the temperature in 2014/2015 and 2000/2002, which
were allowed to vary independently. We find that the temperatures are
virtually identical; at 90\% confidence there is no observable
temperature variation greater than 0.9\% between the 2000/2002 and
2014/2015 epochs. Testing the normalization (which would correspond to
any change in the emitting area) instead, we place an upper limit of
4\% to any change in normalization at 90\% confidence. This is an
indication that the observed radiation from X7 is indeed due to
passive transport of the heat deposited in the stellar core during
outbursts rather than continuing accretion.

In contrast, the thermal X-ray flux of X5 is highly variable on short
timescales (minutes to hours). However, this variability appears not
to be due to on-going accretion but rather due to X5 being an edge-on
system. This geometric configuration results in occultations of the
neutron star and variable absorption by the gas associated with the
residual accretion disk.  As detailed in \S4, this necessitates the
excision of a large fraction of the data for X5 in order to recover
the intrinsic steady flux from the neutron star. Additionally, due to
disk precession, the NS was completely obscured during the deep 2002
exposures, making it difficult to fully assess the long term
variability of the thermal component for X5. These difficulties make
X5 results less reliable than those for X7.

\begin{figure}
  \includegraphics[angle=0,width=0.48\textwidth]{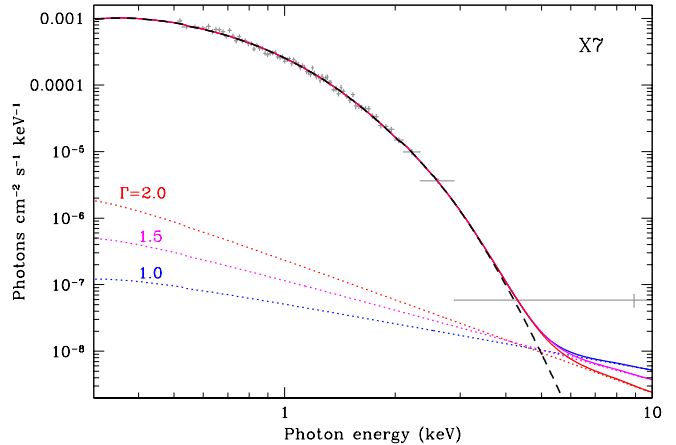}
\caption{The best fit models to the spectrum of X7 with the addition
  of a power-law component (dotted lines) with index $1.0$ (blue),
  $1.5$ (magenta), and $2.0$ (red). The dashed line shows the best-fit
  model ({\tt nsatmos}) for purely thermal hydrogen atmosphere
  emission. For reference, the unfolded \textit{Chandra} sub-array
  data of X7 fitted with a composite model of a H atmosphere plus
  power-law with index 1.5 is show in light grey.}
\label{fig:x7_pl}
\end{figure}

\subsection{Additional Spectral Components}
Another usual cause for concern for the measurements of the neutron
star radius from spectra is the possibility of additional, fainter
X-ray emission components that have not been taken into account in the
spectral modeling. In several qLMXBs, the presence of a power-law
component in the quiescent flux suggests on-going accretion at a low
rate\footnote{The best evidence for accretion at low luminosities in
  neutron star low-mass X-ray binaries comes from the recent detection
  of accretion-powered pulsations with power-law spectra at quiescent
  levels ($\sim$$10^{33}$~erg~s$^{-1}$) in the nearby millisecond
  pulsars PSR~J1023+0038 and XSS~J12270--4859 \citep{Arch15,Pap15}.}
\citep[e.g.,][]{Garcia01}.  Alternatively, ``contaminating'' power-law
emission may arise due to a rotation-powered pulsar wind turning on in
quiescence \citep[see, e.g.,][]{Camp98,Jonk04} or unidentified blended
sources in the crowded globular cluster core.  It is possible that
such a component is present at a very low level in the seemingly
purely thermal qLMXBs but may still skew the NS radius measurement
significantly if it is not accounted for. In particular, the presence
of a power-law component would harden the spectrum so that a fit with
a purely thermal model would produce a higher temperature and,
therefore, $M-R$ limits that are shifted towards lower radii.

To test this scenario, we introduced an additional power-law component
into the model fits. Figure~1 illustrates the maximum possible
contribution of a power-law component to the model spectrum of X7 for
photon indices in the range typical for quiescent LMXB power-laws,
$\Gamma=1.0-2.0$ \citep[][]{Camp98,Cack10,Chak14}.  The addition of a
power-law results in a difference of $0.1$\% in the inferred nominal
NS radius and its associated uncertainties, while the best fits with
and without this component differ in $\chi^2$ by a statistically
insignificant 0.03. We found a similar result for X5, where the
introduction of a power-law causes a change of only 0.5\% in the NS
radius confidence limits.  For X7 and X5, we also derive limits of
$\le$0.2\% and $\le$1.6\%, respectively, for any power-law component
in the subarray spectra. We thus conclude that power-law emission is
negligible for both of these qLMXBs and do not further consider a
power-law component in the spectroscopic analyses that follow.

\subsection{Chemical Composition of the Atmosphere}
Given their transiently accreting nature, the thin atmospheric layer
on the neutron stars found in qLMXBs most likely consists of material
that has the composition of their companion star. For a hydrogen rich
donor, due to gravitational settling, hydrogen is expected to surface
within seconds and thus dominate the surface emission
\citep{Alcock80,Ham83,Brown02}. 

For X5, the measured 8.7-hour orbit \citep{Heinke03} implies that the
donor star is hydrogen-rich and, therefore, the neutron star almost
certainly has a pure hydrogen atmosphere. In the case of X7, due to
the lack of any information regarding the orbital period or the
properties of the secondary star, the chemical composition of the
accreted material is less certain.  For example, this system may be an
ultra-compact binary (with an orbital period $\lesssim$80 min), in
which case the companion star may be a helium star or a C-O core white
dwarf that has surface layers (mostly) devoid of hydrogen \citep[see,
  e.g.,][]{Nel10}. In light of this possibility, \citet{Serv12},
\citet{Cat13}, and \citet{Heinke14} have considered He atmosphere fits
to qLMXB spectra when there is no information about the nature of the
companion. These fits produce significantly larger inferred NS masses
and radii compared to H atmosphere models, owing to the larger
difference between the effective and color temperatures for He
atmosphere emission. These results highlight the importance of the
chemical composition of the NS surface layer on the NS $M$ and $R$
measurements.

\begin{figure}
  \includegraphics[angle=0,width=0.48\textwidth]{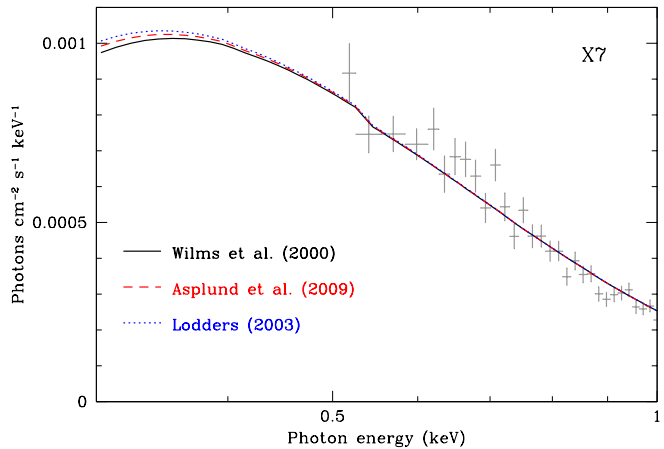}
\caption{The best-fit absorbed hydrogen atmosphere ({\tt nsatmos})
  spectral model for X7 with three choices of interstellar absorption
  model: \citet{Wilms00}, \citet{Aspl09}, and \citet{Lodd03}. Note the
  linear scale on the ordinate for the photon flux. The unfolded
  subarray data for X7 (light grey) fitted with an absorbed {\tt
    nsatmos} model assuming Wilms abundances is shown for reference.}
\label{fig:spec_ism}
\end{figure}

Even in the case of an ultracompact binary and a hydrogen-poor
companion, the question remains as to whether even trace amounts of
hydrogen in the donor are sufficient to cover the neutron star surface
with hydrogen. An optical depth of $\sim$unity can be achieved with a
layer of hydrogen of thickness $\sim$1 cm on the neutron star surface,
which requires only $\sim$$10^{-20}$ M$_{\odot}$ of H (see, e.g.,
\citealt{Zavlin02} and Eq.~(20) of \citealt{Oz13}). As a result, even
in the case of an ultracompact binary with accretion from a He-rich
donor, minute abundances of H can still result in a hydrogen dominated
atmosphere.  An alternative means of accumulating a hydrogen
atmosphere may be through nuclear spallation reactions of nuclei
heavier than helium.  However, it is not certain whether spallation
always produces H during accretion \citep{Bild92,Bild93}.  On the
other hand, H can be depleted from the photospheric layer via
diffusive nuclear burning by an underlying layer of nuclei that are
able to capture protons \citep[see, e.g.,][]{Chang04,Chang10}.  Due to this
ambiguity regarding the atmospheric composition, in \S4 we present fits
with both H and He atmospheres for X7.

\subsection{Interstellar Absorption}
Photoelectric absorption by the interstellar medium along the line of
sight to the qLMXB significantly alters the intrinsic thermal spectrum
of the NS, especially in the very soft X-ray band. Thus, a reliable
constraint on the hydrogen column density along the line of sight,
$N_{\rm H}$, is important for robust NS EoS constraints \citep[see,
  e.g.,][and references therein]{Lat14}. In addition to the total
column density, the observed shape of the thermal spectrum is also
sensitive to the chemical abundances of the intervening material. As a
consequence, the inferred $M-R$ constraints may differ based on the
assumed ISM abundances. This is especially true for targets with large
values of column density $N_{\rm H}$ since the absorption edges due to
metals (whose depth depends on the relative abundances) become much
more prominent.

As demonstrated by \citet{Heinke14}, the strong sensitivity of the
inferred NS radius on the choice of ISM abundances implies that
low-$N_{\rm H}$ targets are best suited for the neutron star EOS
constraints. In addition, the latest abundances and most complete
absorption models that are appropriate for the ISM should be used in
the spectral analyses. The currently best available abundances for the
ISM (as opposed to values derived from the solar spectrum) is that of
\citet{Wilms00} ({\tt wilm} in XSPEC, incorporated into the absorption
model {\it tbabs}), which we use in our analysis. Nevertheless, to
test the sensitivity of the results on the absorption model, we also
repeated the spectroscopic fits by implementing two other abundance
models available in XSPEC: {\tt aspl} \citep{Aspl09} and {\tt lodd}
\citep{Lodd03}, both of which are based on solar abundances. As is
evident from Figure~2, the choice of absorption model does not alter
the model spectral shape above $\sim$0.5~keV due to the exceptionally
low absorbing column towards 47~Tuc ($N_{\rm H} = 1.3\times10^{20}$
cm$^{-2}$).  Therefore, using different abundance models has virtually
no effect on the derived $M-R$ relation. Specifically, for a fixed
$M=1.4$ M$_{\odot}$, the best fit value and associated uncertainties
of $R$ differ by less than $0.6$\% among the three abundance
models. As shown below, for both X7 and X5 this level of uncertainty
introduced by the choice of abundance model is dwarfed by the
uncertainties due to instrument calibration.

\begin{figure}
  \includegraphics[angle=0,width=0.48\textwidth]{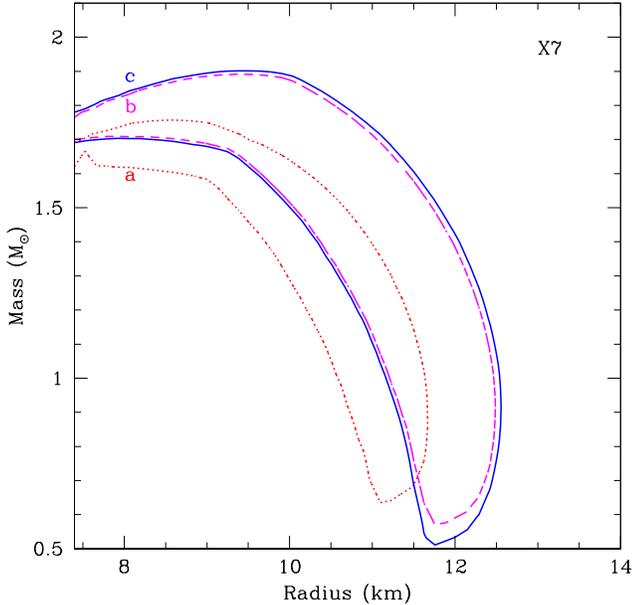}
\caption{68\% confidence contours for the neutron-star mass and radius
  obtained from fitting the spectrum of X7 with an absorbed H
  atmosphere model ({\tt nsatmos}) with three different assumptions in
  the model: a) no pile-up (red dotted line), b) with pile-up (magenta
  dashed line) and c) with pile-up and an additional 3\% systematic
  uncertainty (blue solid line). Note the substantial displacement and
  enlargement of the contours when pile-up is introduced into the
  model.}
\end{figure}

\subsection{Distance Uncertainty}
Another important source of uncertainty in the radius measurements is
the distance to the target qLMXB, which scales linearly with the NS
radius.  Globular cluster distances have been constrained using a
variety of different methods, each of which has its own statistical
and systematic errors (which may not always be well-characterized).
Fortunately, 47~Tuc has been a target for multiple distance
investigations, giving us the opportunity to compare the results from
multiple methods. In addition, the reddening to 47~Tuc is very small
and well measured, nearly eliminating the concern about degeneracy
between distance and reddening with measuring globular cluster
distances\footnote{As discussed in \citet{Heinke14}, no individual
  method, not even cluster kinematics, is entirely free of systematic
  uncertainties.}.

\begin{figure}
  \includegraphics[angle=0,width=0.48\textwidth]{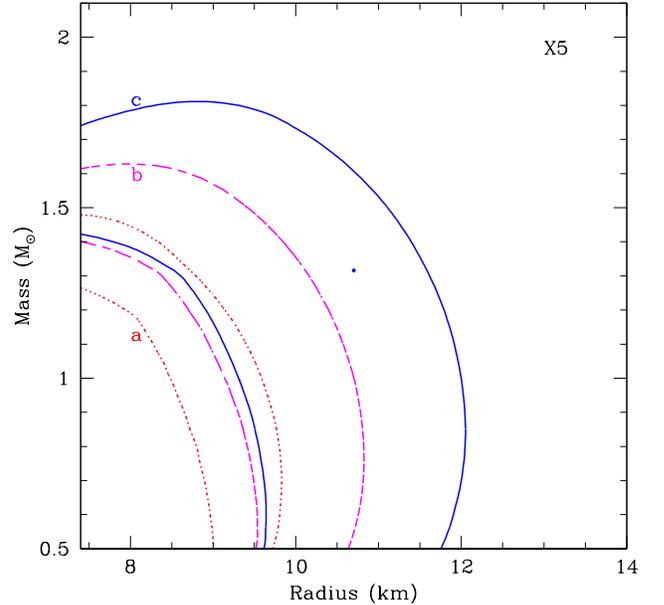}
\caption{68\% confidence contours for the two parameters of interest,
  $M$ and $R$, obtained from fitting the spectrum of X5 with an
  absorbed H atmosphere model ({\tt nsatmos}) with different
  assumptions in the data selection and model: a) only removing the
  eclipses and assuming no pile-up (dotted red line), b) with eclipses
  removed and including pile-up but with no count rate cuts (magenta
  dashed line) and c) with eclipses removed, pile-up included, and
  count rate cuts (solid blue line).}
\end{figure}

A compilation of 22 distance measurements to 47~Tuc, using seven
general methods, is given in Table~1 of \citet{Woodley12}. The methods
include the brightness of the horizontal branch, fitting an isochrone
to the main sequence, cluster kinematics, RR Lyrae stars, the tip of
the red giant branch, eclipsing binaries, and white dwarfs. We add to
this list the recent white dwarf cooling sequence measurement of
\citet{Hansen13}, which reports a distance modulus of
$(m-M)_0 = 13.32 \pm 0.09$ magnitudes. Following \citet{Woodley12}, we
calculated the weighted mean of all these estimates and the error in
this mean to obtain a distance modulus of $(m-M)_0 = 13.31 \pm 0.02$,
corresponding to a distance of $d = 4.59 \pm 0.04$~kpc. For this
calculation, (i) we assumed errors of 0.20 magnitudes for literature
estimates without errors, (ii) we added statistical and systematic
errors in quadrature, and (iii) chose the estimate with binary
corrections from the main-sequence fitting analyses of
\citet{Gratton03} and \citet{Carretta00}.

\begin{figure*}[t!]
  \centering
  \includegraphics[width=0.6\textwidth]{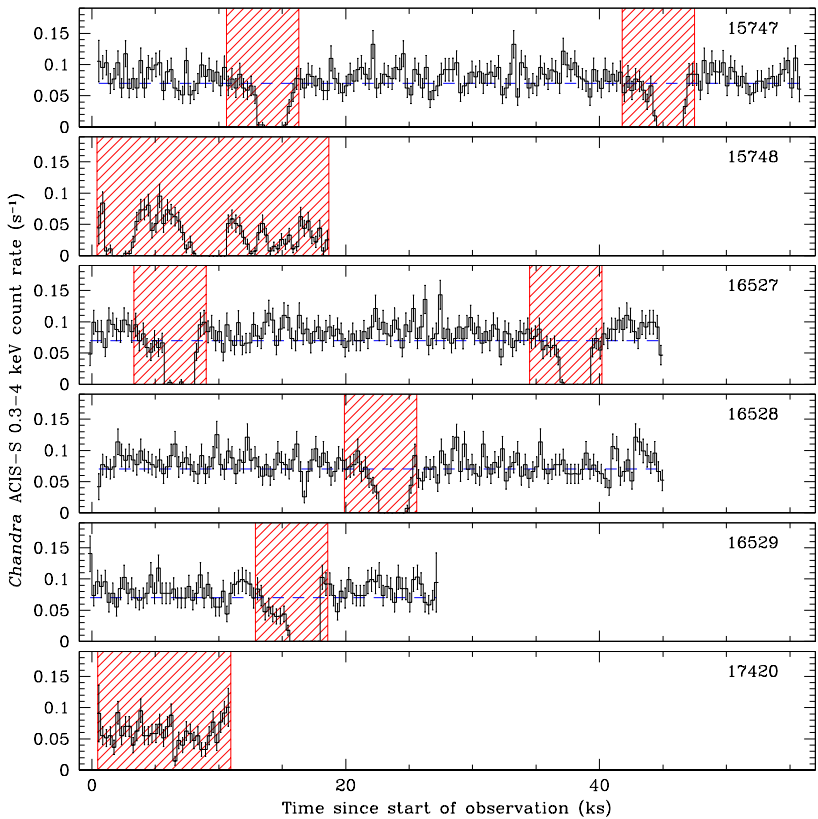}
\caption{The \textit{Chandra} ACIS-S subarray light curves of X5 in
  the 0.3$-$8~keV band. The hatched segments mark the time intervals
  excluded from the spectroscopic analysis due to eclipses or strong
  variability, while the horizontal dashed line marks the count rate
  cut, below which the remaining data were excised.
\label{fig:x5lc}}
\end{figure*}

This analysis has several weaknesses. First, multiple measurements
using the same method may lead to smaller statistical uncertainties
but still contain a bias. Second, improvements in a method over time
may not be reflected in the final calculation if all measurements are
assigned equal weights. Finally, the final measurement could be
significantly affected by results from a single flawed method.  To
address these, we performed a test by selecting the subset of more
recent distance measurements, i.e., those carried out since 2000. We
also undertook ``jackknife'' analyses, where we removed all
measurements taken with one method to assess how much the final result
changes and used the results of this exercise as another measure of
the systematic error. When using the measurements since 2000, we find
a distance modulus of $(m-M)_0 = 13.28 \pm 0.02$ or a distance of $d
=4.53 \pm 0.04$~kpc, which is consistent with the distance measurement
from the full sample.  Our jackknife tests find a range of distance
moduli from $13.27$ to $13.31$, which are also consistent with the
results from the full sample.  Combining the jackknife errors with the
error in the mean in quadrature, we arrive at a final range in
distance modulus of $13.26-13.32$ and, thus, a final distance
measurement of $d = 4.53^{+0.08}_{-0.04}$~kpc.

We note that we excluded from our distance analysis the most recent
measurement of \citet{Wat15}, who obtained a significantly smaller
dynamical distance for 47~Tuc ($d=4.15\pm 0.08$~kpc), compared to the
mean value derived above. This is because we believe that this
reported value is affected by problems in some of the radial velocity
data used in that study. Dynamical distances are derived by comparing
angular proper motions on the sky (typically from multiple
\textit{Hubble Space Telescope} epochs) with radial velocity
dispersions of bright stars.  The radial velocity measurements carried
out by \citet{McLaugh06} and \citet{Lane10} were of individual stars,
but these studies did not check whether multiple stars might be
blended together \citep[as attempted by][]{Geb95}.  Such blending
tends to depress the line-of-sight velocity dispersion in the central
regions.  Indeed, such blending within the central $\sim$1$'$ can be
verified by comparing the positions of stars used for radial
velocities in 47~Tuc by \citet{McLaugh06} with the \textit{HST} image
provided in the same paper, and the radial velocity measurements
within this central region show more scatter than elsewhere.
\citet{Wat15} only used velocity dispersion information from the
central regions (within $<$100$''$) for their primary analysis, to
ensure that proper motions and radial velocity dispersions were
compared within the same region. However, in an Appendix,
\citet{Wat15} show that including velocity dispersion information from
the outer regions leads to a larger inferred distance of
$4.61\pm0.08$~kpc. Thus, we have a plausible explanation for the
discrepancy between the fiducial distance reported by \citet{Wat15},
and the larger distance supported by photometric methods, and by
the consideration of a larger radial velocity database.

\subsection{Instrument Calibration Uncertainties}
The spectroscopic $M-R$ measurement technique using qLMXBs discussed
herein relies on an absolute determination of the flux emitted from
the neutron star. Because of this, it depends strongly on the
reliability of the calibration of the instrument used for the
measurement.  Knowledge of the absolute effective area of the
\textit{Chandra} ACIS detectors is limited by a combination of the
uncertainties in the quantum efficiency near the read out, the quantum
efficiency non-uniformity across the detector resulting from charge
transfer inefficiencies, and the depth of the contaminant on the ACIS
filter (important primarily below $\sim$2 keV).

Following \citet{Gui13}, we adopt a 3\% systematic error to account
for the instrument response uncertainties. We note that an in-depth
evaluation of \textit{Chandra} calibration uncertainties in the
context of qLMXB NS $M-R$ measurements based on the prescriptions by
\citet{Drake06}, \citet{Lee11}, and \citet{Xu14} will be presented in
a subsequent publication.

\begin{figure}
  \includegraphics[width=0.47\textwidth]{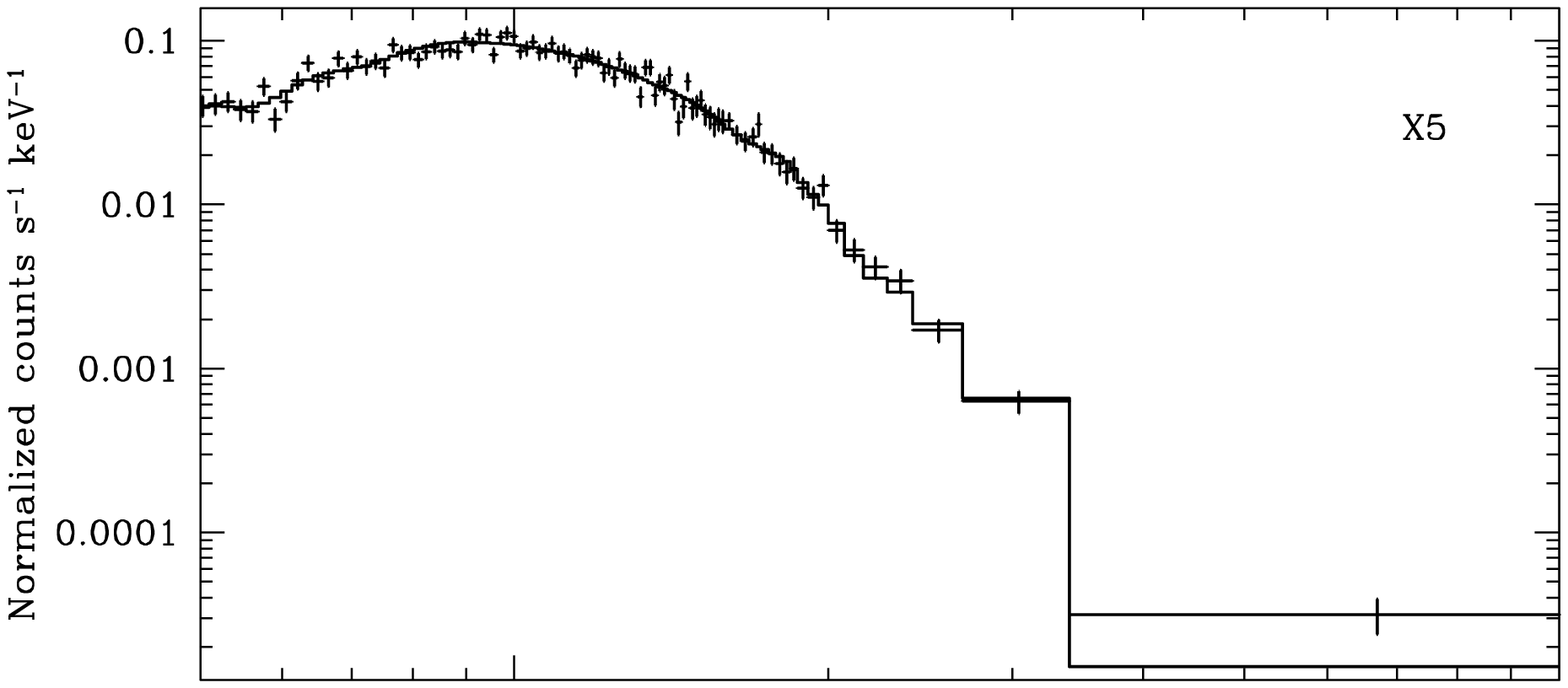}
  \includegraphics[width=0.47\textwidth]{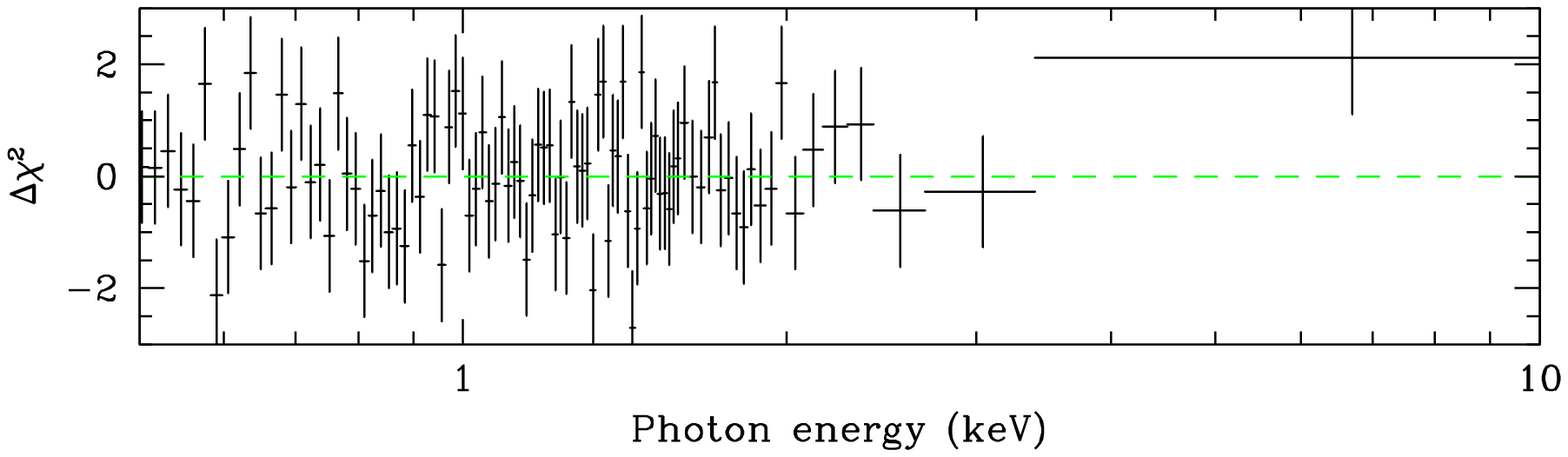}
\caption{The total \textit{Chandra} ACIS-S subarray spectrum of X5
  fitted with an absorbed H atmosphere model ({\tt nsatmos}) convolved
  with a pile-up model (top) and the best-fit residuals (bottom).}
\label{fig:X5_spec}
\end{figure}

\begin{figure}
  \includegraphics[width=0.48\textwidth]{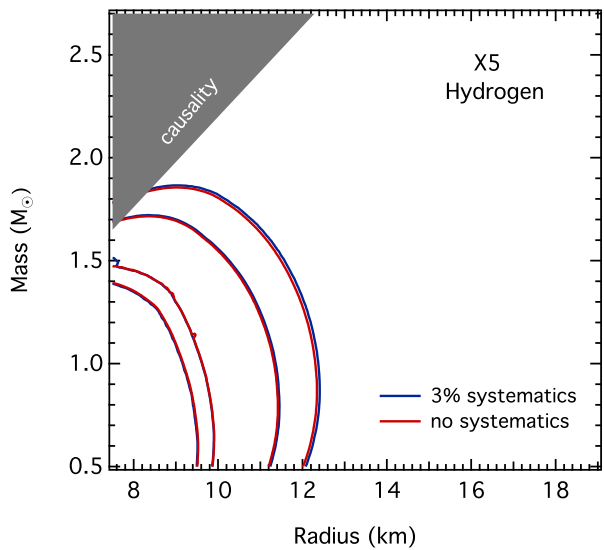}
\caption{The mass-radius constraints obtained for X5 by fitting the
  \textit{Chandra} ACIS-S 1/8 subarray data with a piled-up and
  absorbed hydrogen atmosphere model ({\tt nsatmos}). The 68\% and
  95\% confidence levels are shown, obtained from the posterior
  likelihood over $M$ and $R$ (see text). The blue and red lines correspond,
  respectively, to the fits with and without a 3\% systematic
  uncertainty instrumental calibration uncertainties. The shaded gray
  area in the upper left marks the region excluded by causality
  constraints. See Table~2 for best fit parameters. \label{fig:X5_MR}}
\end{figure}

\begin{figure}
  \includegraphics[width=0.47\textwidth]{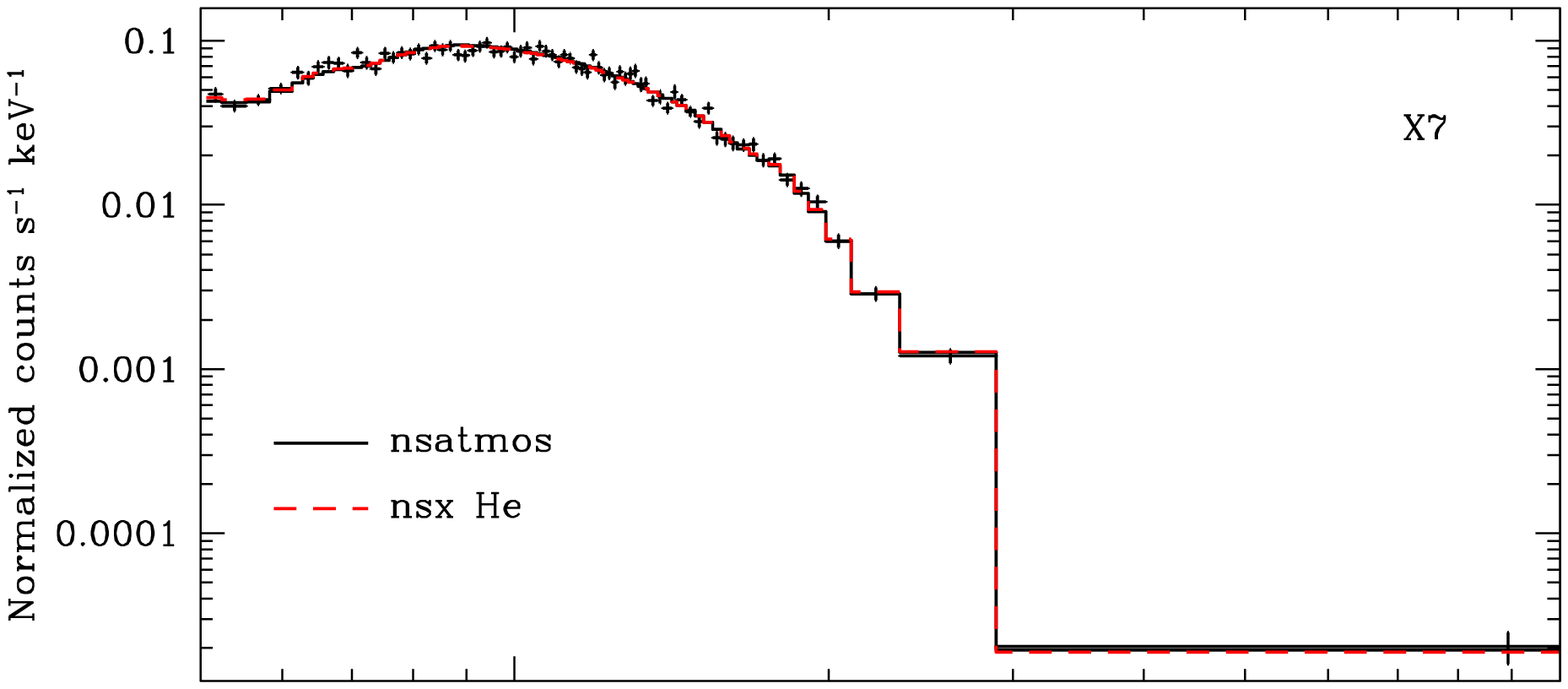}
  \includegraphics[width=0.47\textwidth]{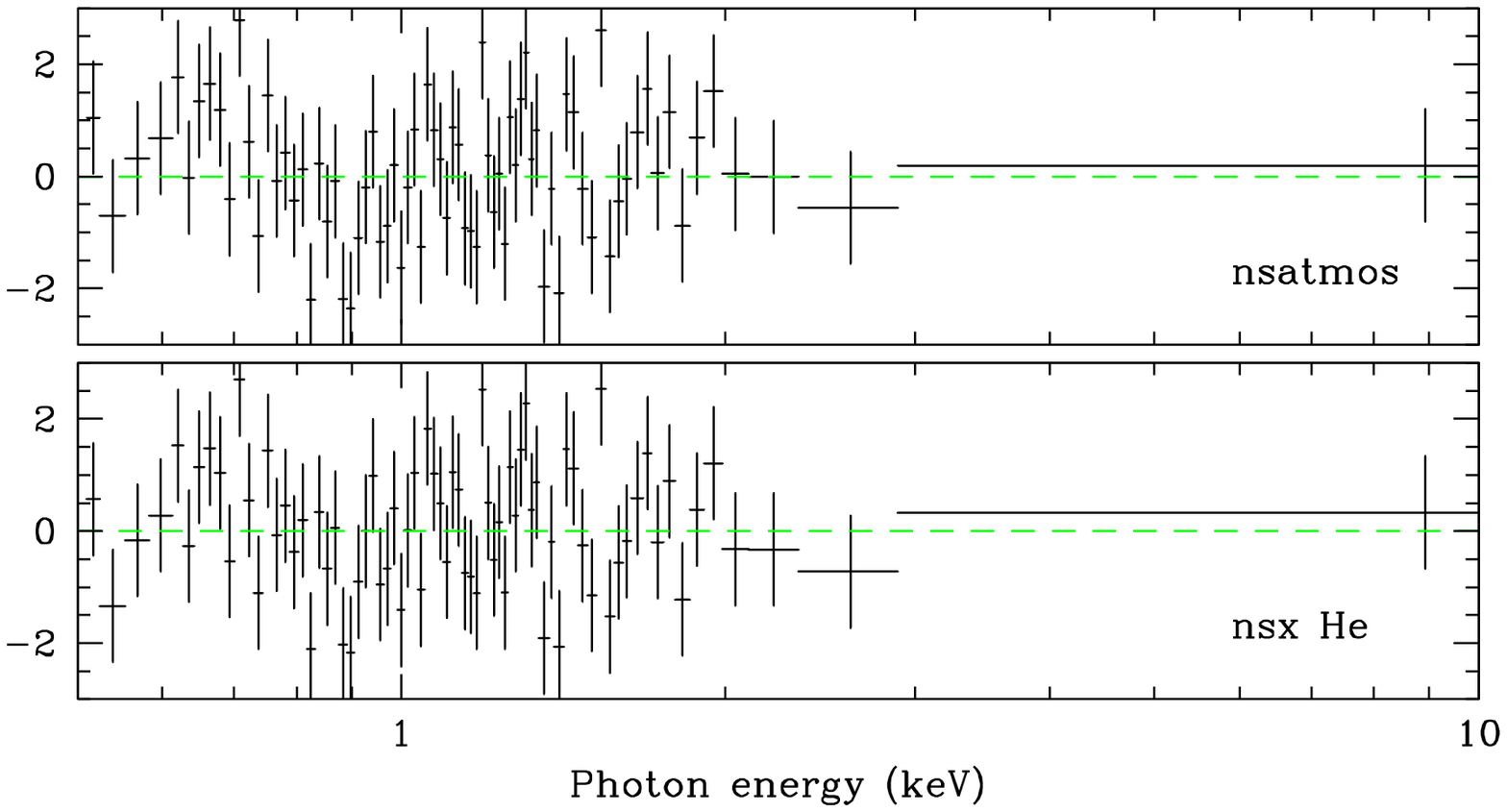}
\caption{The total \textit{Chandra} ACIS-S subarray spectrum of X7
  fitted with an absorbed atmosphere model convolved with a pile-up
  model. The bottom panel shows the best fit residuals expressed in
  terms of $\sigma$. \label{fig:x7_spec}}
\end{figure}

\subsection{Photon Pile-up}
As noted previously, \textit{Chandra} ACIS data of even moderately
bright sources is susceptible to severe event pile-up owing to the
combination of slow readout and high count rate. While we mitigated
most of this negative effect in the new subarray observations by using
a faster detector readout mode, it still affects the data at a low
level. The spectroscopic mass-radius measurement technique for qLMXBs
is highly sensitive to the shape of the thermal spectrum. As a
consequence, even a small artifical distortion in the spectral shape
can bias the $M-R$ measurement. For pile-up specifically, a portion of
the photons piled at lower energies are either rejected entirely (if
their event grades are consistent with those of cosmic rays) or
recorded as a singe photon displaced to higher energies. As reported
in \S2, this occurs for $\sim$1\% of the photons in the X7 and X5
spectra, which we would naively expect to be negligible.

To assess the impact of this seemingly small effect on our results, we
repeated our spectroscopic fits with and without a pile-up model
component (i.e., {\tt pileup} in XSPEC). Figure~3 illustrates the
results for X7, showing the 68\% confidence contours in the $M-R$
plane with and without pile-up. Despite the small degree of pile-up,
the impact on the results is substantial. In addition to enlarging the
confidence intervals, correcting for pile-up displaces them towards
somewhat larger $M$ and $R$. This arises because photon pile-up
artificially hardens the intrinsic source spectrum, which produces a
higher best-fit temperature and hence a smaller inferred stellar
radius. The enlargement of the confidence contours arises principally
from the statistical uncertainty in the pile-up parameter $\alpha$,
which gives the probability of rejection of piled events. Some of the
displacement in the contours may also be due to adding another
parameter to the fit when applying pile-up correction. In light of
these findings, to ensure robust constraints on the NS $M$ and $R$,
pile-up correction should be incorporated into spectral modeling even
for small ($\sim$1-2\%) pile-up fractions.

\section{Results of The Spectroscopic Analyses}

Having explored the various sources of formal and systematic errors,
we performed the optimal spectroscopic fits in XSPEC using the {\tt
  nsatmos} H atmosphere model \citep{Heinke06} for both X5 and X7 as
well as the non-magnetic He atmosphere variant of the {\tt nsx}
additive table model \citep{Ho09} for X7.  We also tested the {\tt
  nsagrav} (Zavlin et al.~1996) H atmosphere model to identify any
discrepancies in the measured radius and mass relative to {\tt
  nsatmos}, which may arise due to differences between the numerical
models used to construct them. The {\tt nsatmos} and {\tt nsagrav}
models produce virtually identical results for the expected range of
combinations of $M$ and $R$. We chose the former in this analysis
since it is defined over a larger range of $M$ and $R$. To account for
interstellar absorption and scattering, we used the {\tt tbabs} model
\citep{Wilms00}.  We further included the correction for the effect of
pile-up, as discussed above. Thus, in what follows, we applied the
multiplicative {\tt pileup} model in XSPEC on the absorbed atmosphere
model under consideration (i.e., {\tt pileup*tbabs*nsatmos}). For the
distance parameter in the {\tt nsatmos} and {\tt nsx} models, we
restricted the allowed range of values to $4.49-4.61$~kpc based on the
discussion presented in \S 3.5. Finally, to incorporate the
instrumental calibration uncertainty, we applied an additional 3\%
measurement error through the {\tt systematic} command in XSPEC.  The
best fit parameters for the various model for both X5 and X7 are
summarized in Table~2.

In order to produce the final confidence contours for X5 and X7 and
use a Bayesian approach to infer the NS EoS parameters from the $M-R$
measurements, we first convert the $\chi^2$ surfaces generated in
XSPEC to a posterior likelihood over $M$ and $R$. For this purpose, we
take advantage of additional information, namely, that no plausible EoS
models predict neutron stars with radii below $\approx$7 km and that
stellar evolution is not expected to produce low mass neutron stars
with $\lesssim$0.5 M$_{\odot}$. Thus, we impose that for $R\le 7$ km
or $M\le 0.5$ M$_{\odot}$, the likelihood goes to zero.

\begin{deluxetable*}{cccccc}
\tablewidth{0pt}
\tablecaption{Summary of mass-radius measurements for X7 and X5}
\tablehead{
\colhead{ } &  \colhead{$N_{\rm H}$\tablenotemark{a}} &  \colhead{$T_{\rm eff}$\tablenotemark{b}} & \colhead{$M$} & \colhead{$R$\tablenotemark{b}} & \colhead{$\chi^2/$d.o.f.}  \\
 \colhead{Model}  & \colhead{($10^{20}$ cm$^{-2}$)}   & \colhead{($10^6$ K)} & \colhead{(M$_{\odot}$)} & \colhead{(km)} & \colhead{}}
\startdata
\multicolumn{6}{c}{X5}\\
\hline
{\tt nsatmos} &  $<$4.9  & $2.74^{+0.21}_{-1.48}$ & $0.5^{+1.27}_{-0}$ & $10.5^{+1.3}_{-10.5}$  & $95.1/88$  \\
              &  $<$5.2  & $2.61^{+0.30}_{-1.23}$ &   ($1.4$) & $9.7^{+1.7}_{-2.0}$  & $95.2/89$  \\
               & $<$5.0 & $1.40^{+1.56}_{-0.15}$ &   $0.84^{+0.62}_{-0.35}$ & ($12$)  & $96.6/89$ \\
\hline
\multicolumn{6}{c}{X7}\\
\hline
{\tt nsatmos} & $<$2.2  & $1.39^{+1.50}_{-0.19}$  & $1.46^{+0.28}_{-1.46}$ & $10.8^{+1.8}_{-10.8}$  & $88.2/70$ \\
              & $<$2.3  & $1.39^{+1.32}_{-0.09}$  &  ($1.4$) & $11.0^{+0.8}_{-0.7}$  & $88.2/71$  \\
              & $<$2.3  & $1.28^{+0.06}_{-0.08}$ &   $1.09^{+0.42}_{-1.09}$ & ($12$)  & $88.7/71$  \\
\hline
{\tt nsx} (He) & $<$3.3 & $1.06^{+1.07}_{-0.06}$  & $0.50^{+1.89}_{-0.50}$ & $14.8^{+2.3}_{-0.6}$  & $82.9/70$  \\
               & $<$3.3 & $1.18^{+0.03}_{-0.07}$  & ($1.4$) & $14.5^{+1.7}_{-0.9}$  & $83.0/71$  \\
               & $<$3.3 & $1.20^{+0.94}_{-0.19}$ & $2.01^{+0.31}_{-0.16}$ & ($12$)  & $83.2/71$
\enddata
\tablenotetext{a}{For $N_{\rm H}$, the lower bound in the fits was fixed to $1.3\times10^{20}$~cm$^{-2}$, 
the value for 47~Tuc. All fits reached this lower limit so only the 90\% confidence upper bound is quoted.}
\tablenotetext{b}{The values quoted are redshift-corrected, i.e., as measured at the neutron star surface.}
\tablenotetext{c}{All quoted uncertainties correspond to 90\% confidence level. Values in parentheses were held 
fixed during the fit.}
\label{resultstable}							 
\end{deluxetable*}

\subsection{X5}

The new $1/8$ subarray \textit{Chandra} ACIS data set reveal that, for
the majority of the time, the accretion disk does not completely
obscure the neutron star in X5. However, during the two shortest
exposures (ObsIDs 15748 and 17420) the system appears highly variable,
possibly due to obscuration by the accretion disk. As already known
from the 2000 and 2005 \textit{Chandra} data, the system undergoes
regular, deep eclipses every $8.666$ hours as well as dips occurring
$\sim$2000~s prior to the main eclipse, which exhibit an enhancement
in $N_{\rm H}$ \citep{Heinke03}. Therefore, it is necessary to
carefully excise the intervals in which X5 suffered eclipses and dips
from our initial spectral analysis. To accomplish this, we extracted
spectra using different time cuts around the eclipses and dips and
fitted the spectra to check for any appreciable changes in the derived
parameters. We show the results in Figure~4.

Throughout the orbit, X5 also exhibits rapid energy-dependent flux
variability, with a spectral hardening at lower count rates. As our
line of sight presumably grazes the accretion disk, this variability
is most likely due to absorption by material from the disk. It is
evident from Figure~4 that the variable nature of X5 can result in
skewed $M-R$ measurements. In light of this, we took great care to
excise the time intervals around the eclipses as well as instances of
strong energy-dependent dimming. We show in Figure~5 the temporal and
count rate cuts we applied to eliminate these portions of the data,
which result in a 88.5~ks effective exposure. We show in Figure~6 the
fitted X-ray continuum and in Figure~7 the resulting 68\% and 95\%
confidence contours for the mass and radius of X5 using the filtered
data. The results indicate a neutron star radius
$R=9.6^{+0.9}_{-1.1}$~km (at 68\% C.L.) for an assumed mass of $M=1.4
\; M_{\odot}$, with $\chi^2_{\nu}=1.07$ for 89 degrees of freedom.  A
radius of 12~km lies outside of the 68\% confidence contour and the
mass is constrained to $M<1.3$ M$_{\odot}$ for that radius at 95\%
confidence level. We note that the lower bound of the 68\% confidence
interval falls below $M = 0.5 \; M_{\odot}$ where the {\tt nsatmos}
model has not been calculated, since stellar evolution suggests that
such low-mass neutron stars are not produced.

\subsection{X7}
We repeated the analysis described above for X7 and show in Figure~8
the total \textit{Chandra} ACIS-S 1/8 subarray spectrum as well as the
best-fit H ({\tt nsatmos}) and He ({\tt nsx}) atmosphere models, which
produce statistically similar fits (with $\chi^2_{\nu}$ of $1.23$ and
$1.17$, respectively). The resulting 68\% and 95\% confidence contours
on the neutron star mass and radius for both models are shown in
Figure~9. Assuming a H atmosphere and for $M=1.4 \; M_{\odot}$, we
obtain a best fit radius $R = 11.1^{+0.8}_{-0.7}$~km. For a He
atmosphere, with the same fixed mass, the fit results in
$R=14.7^{+1.3}_{-0.9}$~km. If we instead hold the radius fixed at
$12$~km, the H atmosphere model results in a low neutron star mass
with a best fit value of $M=1.1^{+0.3}_{-0.4} \; ~M_{\odot}$, while
the He model favors a massive neutron star with $M=2.1^{+0.2}_{-0.2}
\; M_{\odot}$.  All uncertainties quoted above are at a 68\%
confidence level.

\begin{figure*}
  \includegraphics[width=0.48\textwidth]{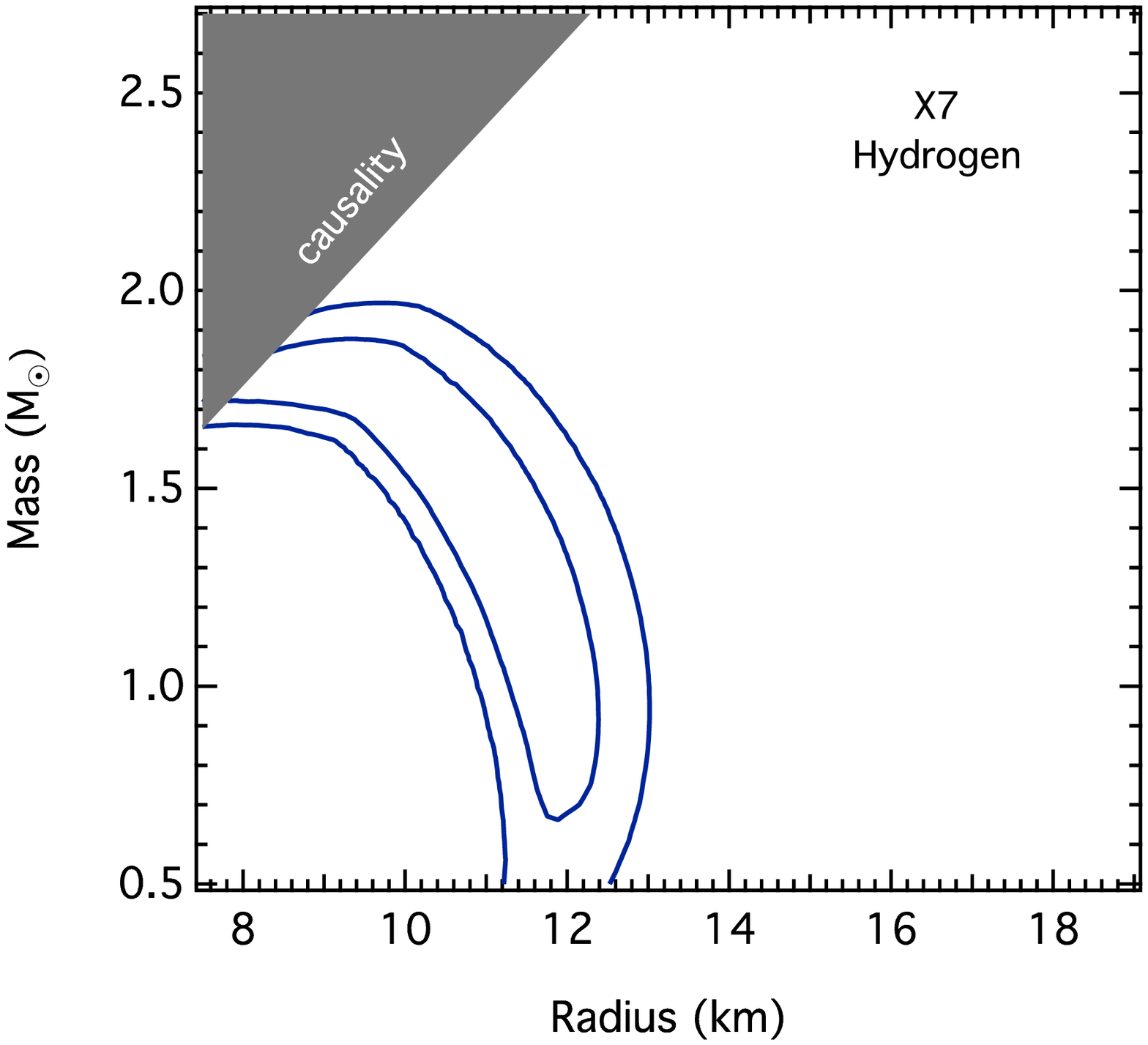}
  \includegraphics[width=0.48\textwidth]{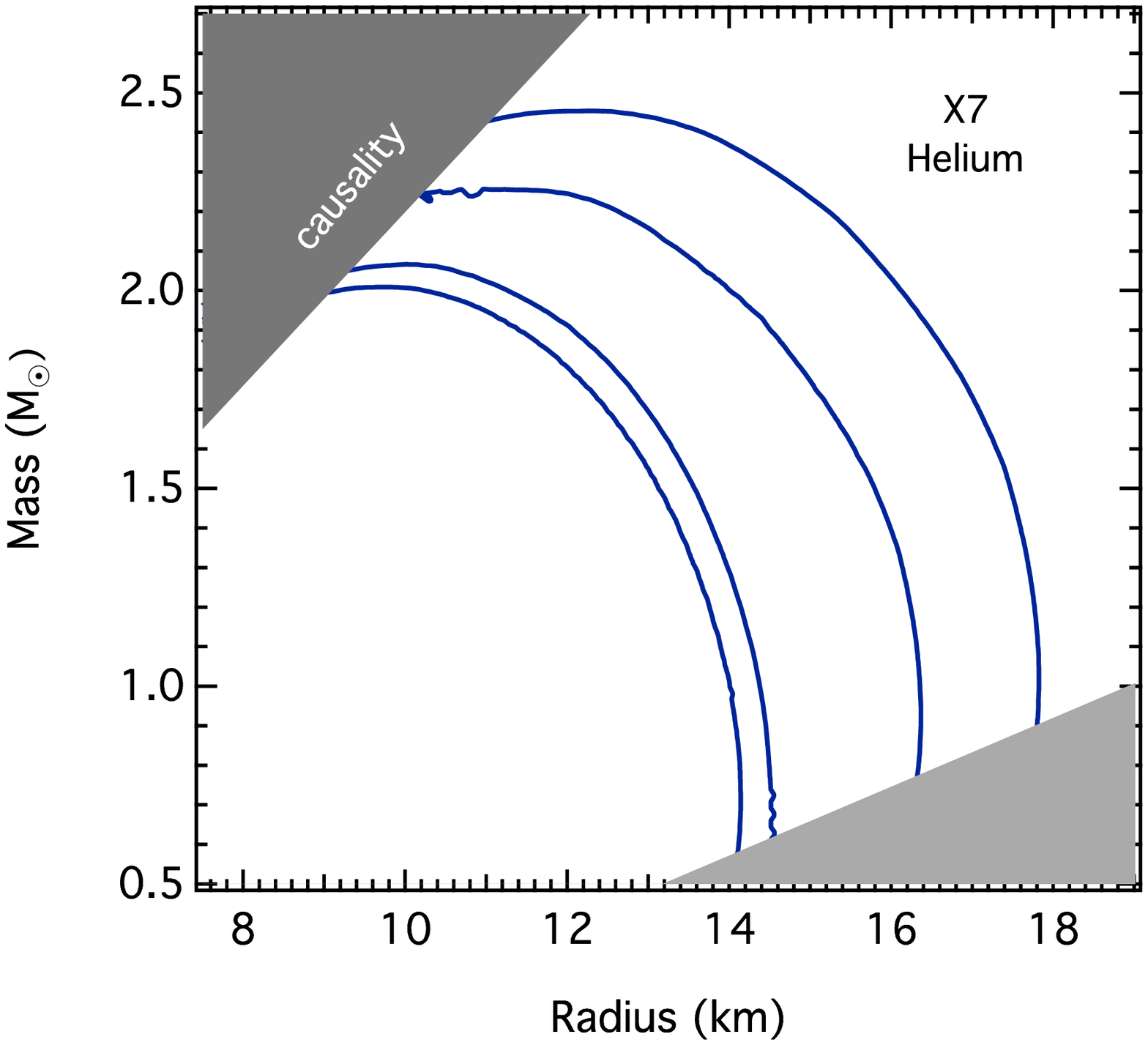}
\caption{The mass-radius constraints obtained for X7 from the
  \textit{Chandra} ACIS-S subarray data assuming the {\tt nsatmos} H
  atmosphere (left) and {\tt nsx} He atmosphere (right) models. 68\%
  and 95\% confidence contours are shown obtained from the posterior
  likelihood over $M$ and $R$ (see text). A 3\% systematic uncertainty in the
  calibration and a model for 1\% pile-up are included in the
  analyses. See Table~2 for the best fit parameters.}
\end{figure*}

\subsubsection{X7: Hydrogen or Helium?}
As noted previously, while there are numerous theoretical reasons to
expect a hydrogen atmosphere, in the absence of any information
regarding the orbital parameters and companion properties of X7, we
cannot definitively determine whether the neutron star atmosphere is
dominated by hydrogen or helium. However, there are some astrophysical
arguments in favor of a lower-mass star and hence a H atmosphere.

Neutrino emission and cooling mechanisms, including direct URCA, hyperon
URCA, and quark core cooling, have strong dependences on properties of the
core, such as density and proton fraction, and, therefore, on the nuclear
equation of state and neutron star mass \citep{Yak04,Bez15}.  The
transiently accreting X-ray binary SAX J1808.4$-$3658 has a very cold NS,
compared to what would be expected through ``standard'' cooling,
whereas most other observed NSs in X-ray binaries are much warmer
\citep[see][]{Heinke07,Heinke09,Wij13}. The low temperature of SAX
J1808.4$-$3658 suggests it has a higher mass than that of other
quiescent NSs, with the latter presumably near the typical mass of 1.4
M$_{\odot}$ and current evidence pointing to SAX J1808.4$-$3658 having
$M_{NS} \lesssim 1.6$ M$_{\odot}$ (at $\sim$2$\sigma$)
\citep{Wang13}. In comparison, the relatively high temperature of X7
suggests it has a comparable, if not lower, mass, i.e., around $1.4$
M$_{\odot}$ or lighter.  This is contrary to our He atmosphere fits
which yield $M\gtrsim$1.7 M$_{\odot}$ (see the right panel of
Figure~9) and would result in a X7 temperature that is
at least as low as that of SAX J1808.4$-$3658, contrary to observations.

We find that a He atmosphere model would require either a quite large
radius or a high mass. A high mass would be at odds with our current
understanding of cooling processes based on other
qLMXBs, while a large radius ($>$14 km) would conflict both with the
measurements of the radius of X5 presented here, and with measurements
of other NSs \citep{Gui11,Heinke14,Oz15}, including the radius derived
for the NS in NGC 6397 for either H or He atmospheres. Based on this
line of reasoning, we argue that a hydrogen composition for the
atmosphere of X7 is more plausible.

\section{Implications for the Neutron Star Equation of State}
Most existing mass-radius measurements of neutron star have
uncertainties that are too large to offer useful constraints when
considered individually.  Nevertheless, as shown recently in
\citet{Oz15}, when taken as an ensemble, these results can produce
fairly tight constraints on the equation of state. We now add the
$M-R$ measurements of X7 and X5 to those used in the earlier study to
infer the NS equation of state.

To accomplish this, we make use of the Bayesian statistical framework
developed in \citet{Oz15} to measure the most likely dense matter
equation of state and the corresponding neutron star radii from all of
the spectroscopic measurements. This framework makes use of parametric
representations of the equation of state and allows us to use these
radius measurements to directly infer the pressures at several
fiducial densities above the nuclear saturation density, i.e., at
$\rho_1=1.85 \; \rho_{\rm sat}$, $\rho_2=3.7 \; \rho_{\rm sat}$, and
$\rho_3= 7.4 \; \rho_{\rm sat}$ by exploiting the unique mapping
between the pressure-density relation of cold supra-nuclear matter and
the neutron star $M-R$ relation
\citep{Lind92,Lat01,OP09,Read09,Heb10}. This statistical inference
also allows us to consider additional information and constraints on
the dense matter equation of state, such as the results of laboratory
experiments in the vicinity of nuclear saturation density
\citep[see][and references therein]{Tsang12,Lat13}, the requirement
for a $\ge$2 M$_{\odot}$ maximum neutron star mass
\citep{Dem10,Ant13},\citep[also, see reviews by][]{Lat11,OF16}, the
physical conditions of stability and causality for the parametric
equation of state, as well as different priors on the pressures $P_1,
P_2$, and $P_3$ at the fiducial densities (see Section~5 of
\citealt{Oz15} for additional details).

\begin{figure*}
  \includegraphics[width=0.48\textwidth]{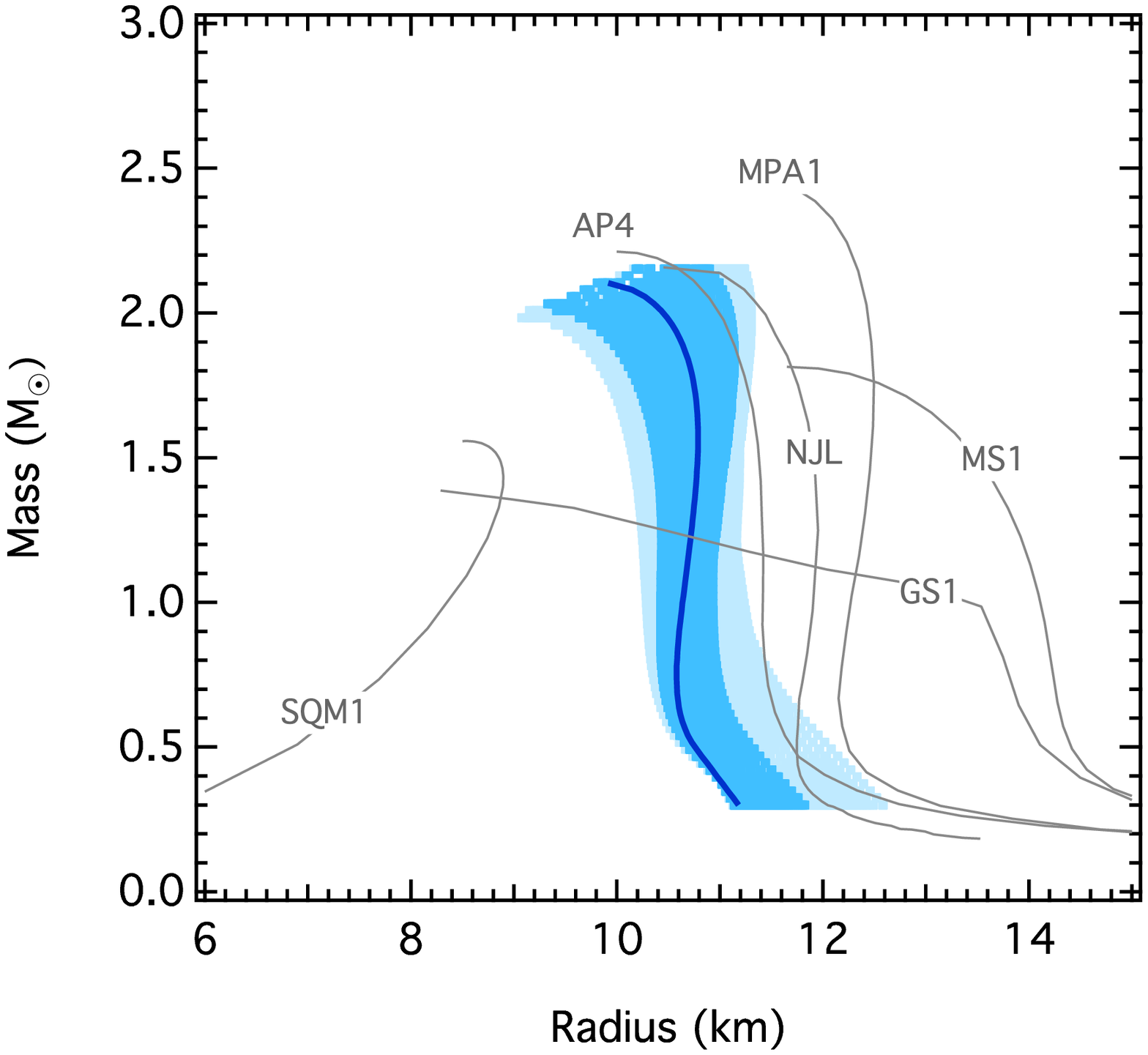}
  \includegraphics[width=0.48\textwidth]{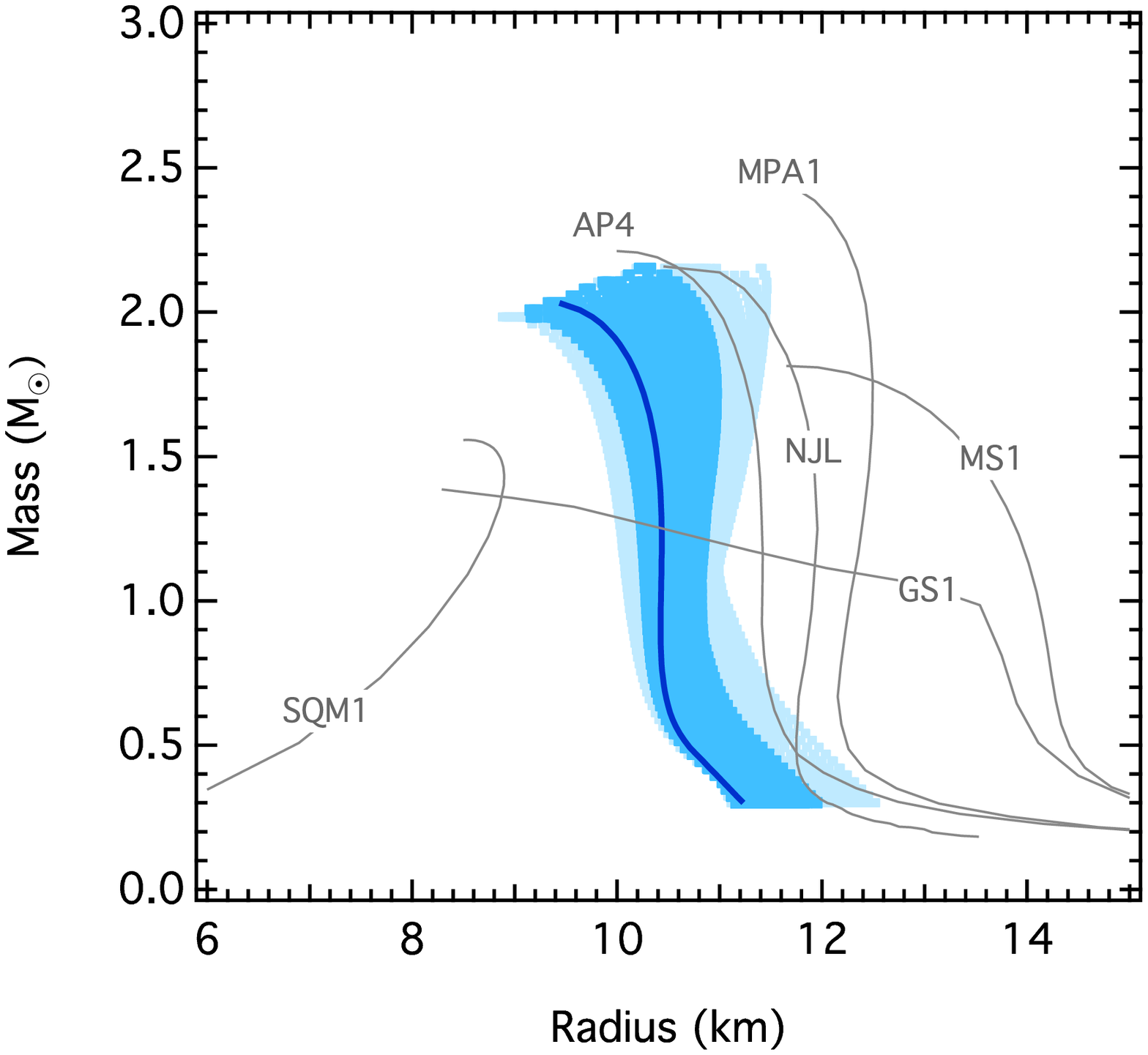}
\caption{The mass-radius relation (solid blue curve) corresponding to
  the most likely triplet of pressures that agrees with the current
  neutron star data. These include the X5 and X7 radius measurements
  shown in this work, as well the neutron star radii measurements for
  the twelve neutron stars included in \citet{Oz15}, the low energy
  nucleon-nucleon scattering data, and the requirement that the EoS
  allow for a $M > 1.97 M_\odot$ neutron star. The ranges of
  mass-radius relations corresponding to the regions of the $(P_1,
  P_2, P_3)$ parameter space in which the likelihood is within
  $e^{-1/2}$ and $e^{-1}$ of its highest value are shown in dark and
  light blue bands, respectively. The results for both flat priors in
  $P_1, P_2,$ and $P_3$ (left panel) and for flat priors the
  logarithms of these pressures (right panel) are shown.}
\label{fig:EOS_const}
\end{figure*}

Using this framework, we combine the radius likelihoods for X7 and X5
presented here with those of the twelve sources included in
\citet{Oz15} to infer the posterior likelihoods over the pressures at
the three fiducial densities. In Figure~10, we show the mass-radius
relation that corresponds to the most likely triplet of pressures that
is derived from the combined likelihoods of the fourteen sources as
well as the priors on the EoS discussed above. Specifically, these
include the hydrogen atmosphere results for X7, based on the arguments
in \S5.1, as well as the six neutron stars for which thermonuclear
burst data have been used to infer $M$ and $R$ and the six qLMXBs
analyzed in \"Ozel et al.~(2015). We also show in Figure~10 the ranges
of mass-radius relations that correspond to the regions of the $(P_1,
P_2, P_3)$ parameter space in which the posterior likelihood is within
$e^{-1/2}$ and $e^{-1}$ of its highest value. For comparison, we
include in this figure a small selection of proposed EoS with
different compositions and calculation techniques (see \citealt{Oz15}
for the acronyms and further details).

It is evident from this figure that the empirical equation of state
that is preferred by the current radius measurements is consistent
with relatively small radii. In particular, around $1.5M_\odot$, the
95\% confidence range spans 9.9 to 11.2~km. This also indicates a
fairly soft EoS; i.e., a lower pressure at and above $2\; \rho_{\rm
  sat}$ than those predicted by a number of nucleonic equations of
state, such as AP4 \citep{Akmal98} shown in this figure.

\section{Conclusions}
We presented spectroscopically derived neutron star mass-radius
constraints for the qLMXBs X5 and X7 in the globular cluster 47 Tuc
based on new \textit{Chandra} observations that were specifically
optimized for this purpose. Although significantly shorter than the
previous 270-ks ACIS-S full-frame exposure of 47 Tuc \citep{Heinke05},
the 181-ks 1/8 subarray data of 47 Tuc we have presented here results
in much tighter constraints on the $M-R$ relation for X7.  This is a
direct consequence of the use of a faster readout mode for the ACIS-S
detector, which significantly reduced the pile-up fraction (from
$\sim$15\% to $\sim$1\%) and, therefore, the distortion of the spectra
due to the effects of pile-up. Furthermore, this substantially reduces
the unquantified systematic uncertainties introduced by pile-up, which
exibits non-linear behavior as a function of count rate and is not
well calibrated. Indeed, the limits on $M$ and $R$ for X7 obtained in
\citet[][see, in particular, their Figure 2]{Heinke06} are only
marginally consistent with those shown in Figure 9 using the same {\tt
  nsatmos} H atmosphere model. This leads us to conclude that qLMXB
data affected by a high pile-up fraction are not useful for reliable NS
EoS constraints.

Another important finding of our evaluation of sources of measurement
uncertainty described in \S3 is that even at a level of $\sim$1\%,
ignoring photon pile-up in the spectroscopic fits can lead to not only
underestimated uncertainties but also a skewed measurement (as
illustrated in Figures 3 and 4). This suggests that all previous qLMXB
analyses that have not applied any necessary pile-up corrections in the
spectroscopic fits should be re-examined. (Note that pile-up is well
below 1\% for a number of \textit{Chandra} observations of qLMXBs and
the previous analyses of the qLMXB in M28 take into account a pile-up
correction for a $\sim 4\%$ pile-up fraction in those data; see
\citealt{Gui13}.)

Unlike most other globular cluster qLMXBs, the observed spectra of X7
and X5 are not strongly attenuated by photoelectric absorption from
interstellar gas. As a result, the spectroscopic constraints we have
obtained here do not suffer from appreciable uncertainties due to lack
of information on the relative chemical abundances of the interstellar
medium \citep[see][, for further details]{Heinke14}.  When taken
together with other factors, such as the well-determined distance to
47 Tuc, this makes the X7 radius measurements and the resulting NS EoS
constraints presented here some of the most robust based on this
spectrosopic technique.

The mass and radius constraints we find for X5 and X7 through these
new observations are highly consistent with the twelve other
spectroscopic radius measurements that have been performed to date for
qLMXBs and thermonuclear bursters (see \citealt{Oz15}). The increase
in the radius measurements as well as the consistency of the results
for different sources with a variety of different uncertainties
increases our confidence in these measurements. Furthermore, it allows
us to place increasingly tighter constraints on the dense matter
equation of state, especially when combined with other measurements,
such as nuclear experiments near nuclear saturation density and the
high neutron star masses measured through pulsar timing.

Indeed, by combining the results for X5 and X7 with existing $M-R$
measurements from other qLMXB as well as bursting neutron stars, we
obtain increasingly more robust constraints on the neutron star
equation of state.  Specifically, we find that the preferred equation
of state that is empirically derived from the measurements of all
fourteen sources predicts radii between 9.9 and 11.2~km around $M=1.5
M_\odot$ (corresponding to the range where the likelihood falls to
$e^{-1}$ of its maximum value). This also implies a relatively low
pressure around twice nuclear saturation density, which most directly
affects neutron star radii. We find that such an equation of state can
easily produce $\sim 2 M_\odot$ neutron stars that is observed through
radio pulsar timing. This preferred equation of state is softer than
some purely nucleonic equations of state that are tuned to fit
experiments at low densities, such as AP4, and may point to new
degrees of freedom appearing around $\sim 2 \; \rho_{\rm sat}$ in
neutron-rich matter.

\acknowledgements We thank W.C.G.~Ho for numerous helpful
comments. This work was funded in part by NASA \textit{Chandra} grants
GO4-15029A and GO4-15029B awarded through Columbia University and the
University of Arizona and issued by the \textit{Chandra} X-ray
Observatory Center, which is operated by the Smithsonian Astrophysical
Observatory for and on behalf of NASA under contract NAS8-03060. COH
acknowledges support from an NSERC Discovery Grant and a Humboldt
Fellowship, and the hospitality of the Max Planck Institute for Radio
Astronomy, Bonn, Germany. TG was supported by Scientific Research
Project Coordination Unit of Istanbul University, Project numbers
49429 and 57321. This research has made extensive use of the NASA
Astrophysics Data System (ADS), the arXiv, and software provided by
the Chandra X-ray Center (CXC) in the application package CIAO.

{\it Facilities:} \facility{CXO (ACIS)}.


\begin{thebibliography}{}

\bibitem[Akmal et al.(1998)]{Akmal98} Akmal, A., Pandharipande, V.~R., Ravenhall, D.~G.~1998, \prc, 58, 1804

\bibitem[Alcock \& Illarionov(1980)]{Alcock80} Alcock, C., \& Illarionov, A.~1980, \apj, 235, 534
  
\bibitem[Antoniadis et al.(2013)]{Ant13} Antoniadis, J., Freire, P.~C.~C., Wex, N., et al.~2013, Science, 340, 448 
  
\bibitem[Archibald et al.(2015)]{Arch15}  Archibald, A.~M., Bogdanov, S., Patruno, A., et al.~2015, \apj, 807, 62

\bibitem[Asplund et al.(2009)]{Aspl09} Asplund, M., Grevesse, N., Sauval, A. J., \& Scott, P.~2009, ARA\&A, 47, 481

\bibitem[Bahramian et al.(2015)]{Bah15} Bahramian, A., Heinke, C.~O., Degenaar, N., et al.~2015, MNRAS, 452, 3475

\bibitem[Baub\"ock et al.(2015)]{Baub15} Baub\"ock, M., \"Ozel, F., Psaltis, D., \& Morsink, S.~M.~2015, \apj, 799, 22
  
\bibitem[Becker et al.(2003)]{Beck03} Becker, W., Swartz, D.~A., Pavlov, G. G.,~et al.~2003, \apj, 594, 798

\bibitem[Beznogov \& Yakovlev(2015)]{Bez15} Beznogov, M.~V., \& Yakovlev, D.~G. 2015, \mnras, 452, 540
  
\bibitem[Bildsten et al.(1992)]{Bild92} Bildsten, L., Salpeter, E.~E., \& Wasserman, I.~1992, \apj 384, 143

\bibitem[Bildsten et al.(1993)]{Bild93} Bildsten, L., Salpeter, E.~E., \& Wasserman, I.~1993, \apj 408, 615
  
\bibitem[Brown et al.(2002)]{Brown02} Brown, E.~F., Bildsten, L., Chang, P.~2002, \apj, 574, 920

\bibitem[Brown et al.(1998)]{Brown98} Brown, E.~F., Bildsten, L., Rutledge, R.~E.~1998, \apj, 504, L95
  
\bibitem[Cackett et al.(2010)]{Cack10} Cackett, E.~M., Brown, E.~F., Miller, J.~M., Wijnands, R.~2010, \apj, 720, 1325

\bibitem[Campana et al.(1998)]{Camp98} Campana, S., Colpi, M., Mereghetti, S., Stella, L., Tavani, M.~1998, A\&Arv, 8, 279 
  
\bibitem[Carretta et al.(2000)]{Carretta00} Carretta, E., Gratton, R.~G., Clementini, G., Fusi Pecci, F.~2000, \apj, 533, 215
  
\bibitem[Catuneanu et al.(2013)]{Cat13} Catuneanu, A., Heinke, C.~O., Sivakoff, G.~R., Ho, W.~C.~G., Servillat, M.~2013, \apj, 764, 145   
  
\bibitem[Chakrabarty et al.(2014)]{Chak14} Chakrabarty, D., Tomsick, J.~A.; Grefenstette, B.~W.,~et al.~2014, \apj, 797, 92 
  
\bibitem[Chang \& Bildsten(2004)]{Chang04} Chang, P., \& Bildsten, L.~2004, \apj, 605, 830

\bibitem[Chang et al.(2010)]{Chang10} Chang, P., Bildsten, L., Arras, P.~2010, \apj, 723, 719
  
\bibitem[Fruscione et al.(2006)]{Fruscione06} Fruscione, A., et al.\ 2006, \procspie, 6270

\bibitem[Davis(2001)]{Davis01} Davis, J.~E.~2001, \apj, 562, 575
  
\bibitem[Demorest et al.(2010)]{Dem10} Demorest, P.~B., Pennucci, T., Ransom, S.~M., Roberts, M.~S.~E., Hessels, J.~W.~T.~2010, Nature, 467, 1081

\bibitem[Drake et al.(2006)]{Drake06} Drake, J.~J., Ratzlaff, P., Kashyap, V., Edgar, R., Izem, R., Jerius, D., Siemiginowska, A., Vikhlinin, A.~2006, SPIE, 6270, 1
  

\bibitem[Garcia et al.(2001)]{Garcia01} Garcia, M.~R., 
McClintock, J.~E., Narayan, R., et al.\ 2001, \apjl, 553, L47

\bibitem[Gebhardt et al.(1995)]{Geb95} Gebhardt, K., Pryor, C., Williams, T.~B., Hesser, J.~E.~1995, AJ, 110, 1699
  
\bibitem[Gendre et al.(2003a)]{Gend03a} Gendre, B., Barret, D., Webb, N.~2003, A\&A, 403, L11
  
\bibitem[Gendre et al.(2003b)]{Gend03b} Gendre, B., Barret, D., Webb, N.~A.~2003, A\&A, 400, 521
  
\bibitem[Gratton et al.(2003)]{Gratton03} Gratton, R.~G., Bragaglia, A., Carretta, E., Clementini, G., Desidera, S., Grundahl, F., \& Lucatello, S.~2003, A\&A, 408, 529
  
\bibitem[Guillot et al.(2011)]{Gui11} Guillot, S., Rutledge, R.~E., Brown, E.~F.~2011, \apj, 732, 88
  
\bibitem[Guillot et al.(2013)]{Gui13} Guillot, S., Servillat, M., Webb, N.~A., Rutledge, R.~E.~2013, \apj, 772, 7

\bibitem[Guillot \& Rutledge(2014)]{Gui14}  Guillot, S., \& Rutledge, R.~E.~2014, \apj, 796, L3
  
\bibitem[Haakonsen et al.~(2012)]{Haa12} Haakonsen, C.~B., Turner, M.~L.; Tacik, N.~A., Rutledge, R.~E.~2012, \apj, 749, 52

\bibitem[Hameury et al.(1983)]{Ham83} Hameury, J.~M., Heyvaerts, J., Bonazzola, S.~1983, A\&A, 121, 259
   
\bibitem[Hansen et al.(2013)]{Hansen13} Hansen, B.~M.~S., Kalirai, J.~S., Anderson, J., et al.~2013, Nature, 500, 51 

\bibitem[Hebeler et al.(2010)]{Heb10} Hebeler, K., Lattimer, J.~M., Pethick, C.~J., Schwenk, A.~2010, Physical Review Letters, 105, 161102 
  
\bibitem[Heinke et al.(2003)]{Heinke03} Heinke, C.~O., Grindlay, J.~E., Lloyd, D.~A., Edmonds, P.~D.~2003, \apj, 598, 501

\bibitem[Heinke et al.(2005)]{Heinke05} Heinke, C.~O., Grindlay, J.~E., Edmonds, P.~D., et al.~2005, \apj, 625, 796
  
\bibitem[Heinke et al.(2006)]{Heinke06} Heinke, C.~O., Rybicki, G.~B., Narayan, R., Grindlay, J.~E.~2006, \apj, 644, 1090

\bibitem[Heinke et al.(2007)]{Heinke07} Heinke, C.~O., Jonker, P.~G., Wijnands, R., Taam, R.~E.~2007, \apj, 660, 1424
  
\bibitem[Heinke et al.(2009)]{Heinke09} Heinke, C.~O., Jonker, P. G., Wijnands, R., Deloye, C.~J., Taam, R.~E.~2009, \apj, 691, 1035
  
\bibitem[Heinke et al.(2014)]{Heinke14} Heinke, C.~O., Cohn, H.~N., Lugger, P.~M., et al.~2014, MNRAS, 444, 443

\bibitem[Ho \& Heinke(2009)]{Ho09} Ho, W.~C.~G., \& Heinke, C.~O.~2009, Nature, 462, 71

\bibitem[Jonker et al.(2004)]{Jonk04} Jonker, P.~G., Galloway, D.~K., McClintock, J.~E., Buxton, M., Garcia, M., Murray, S.~2004, \mnras, 354, 666

\bibitem[Lane et al.(2010)]{Lane10} Lane, R.~R., Kiss, L.~L., Lewis, G.~F., Ibata, R.~A., Siebert, A., Bedding, T.~R., Sz{\'e}kely, P.~2010, \mnras, 401, 2521

\bibitem[Lattimer \& Lim(2013)]{Lat13} Lattimer, J.~M., \& Lim, Y.~2013, \apj, 771, 51

\bibitem[Lattimer(2011)]{Lat11} Lattimer, J.~M.~2011, \apss, 336, 67
  
\bibitem[Lattimer \& Prakash(2001)]{Lat01} Lattimer, J.~M., \& Prakash, M.~2001, \apj, 550, 426
  
\bibitem[Lattimer \& Prakash(2007)]{Lat07} Lattimer, J.~M., \& Prakash, M.~2007, PhR, 442, 109
  
\bibitem[Lattimer \& Steiner(2014)]{Lat14} Lattimer, J.~M., \& Steiner, A.~W.~2014, \apj, 784, 123

\bibitem[Lee et al.(2011)]{Lee11} Lee, H., Kashyap, V.~L., van Dyk, D.~A., et al.~2011, \apj, 731 126

\bibitem[Lindblom(1992)]{Lind92} Lindblom, L.~1992, \apj, 398, 569
  
\bibitem[Lodders(2003)]{Lodd03} Lodders, K.~2003, \apj, 591, 1220

\bibitem[Lugger et al.(2007)]{Lug07} Lugger, P.~M., Cohn, H.~N., Heinke, C.~O., Grindlay, J.~E., Edmonds, P.~D.~2007, \apj, 657, 286

\bibitem[McLaughlin et al.(2006)]{McLaugh06} McLaughlin, D.~E., Anderson, J., Meylan, G., Gebhardt, K., Pryor, C., Minniti, D., Phinney, S.~2006, ApJS, 166, 249
  
\bibitem[Nelemans \& Jonker(2010)]{Nel10} Nelemans, G., \& Jonker, P. G.~2010, NewAR, 54, 87  

\bibitem[{\"O}zel(2013)]{Oz13} {\"O}zel, F.\ 2013, Reports on Progress
  in Physics, 76, 016901

\bibitem[\"Ozel et al.(2010)]{Oz10} \"Ozel, F., Baym, G., \& G\"uver, T.~2010, Phys. Rev. D, 82, 101301

\bibitem[\"Ozel \& Freire(2016)]{OF16} \"Ozel, F. \& Freire, P.\ 2016, Ann. Rev. Astron. Astrophys., in press

\bibitem[{\"O}zel \& Psaltis(2009)]{OP09} {\"O}zel, F., \& Psaltis, D.\ 2009, \prd, 80, 103003 

\bibitem[\"Ozel et al.(2015)]{Oz15} Ozel, F., Psaltis, D., Guver, T., Baym, G., Heinke, C., Guillot, S.~2015, \apj,  in press (eprint arXiv:1505.05155)
 
\bibitem[Papitto et al.(2015)]{Pap15} Papitto, A., de Martino, D., Belloni, T.~M., et al.~2015, MNRAS, 449, L26


\bibitem[Potekhin(2014)]{Pot14} Potekhin, A.~Yu~2014, Physics Uspekhi,, 57, 735
  
\bibitem[Rajagopal \& Romani(1996)]{Raj96} Rajagopal, M., \& Romani, R.~W.~1996, \apj, 461, 327

\bibitem[Read et al.(2009)]{Read09} Read, J.~S., Lackey, B.~D., Owen, B.~J., Friedman, J. L.~2009, \prd, 79, 124032
  
\bibitem[Rutledge et al.(1999)]{Rut99} Rutledge, R.~E., Bildsten, L., Brown, E.~F., Pavlov, G.~G., Zavlin, V.~E.~1999, \apj, 514, 945

\bibitem[Rutledge et al.(2001a)]{Rut01a} Rutledge, R.~E., Bildsten, L., Brown, E.~F., Pavlov, G.~G., Zavlin, V.~E. et al.~2001a, \apj, 551, 921

\bibitem[Rutledge et al.(2001b)]{Rut01b} Rutledge, R.~E., Bildsten, L., Brown, E.~F., Pavlov, G.~G., Zavlin, V.~E. et al.~2001b, \apj, 559, 1054  

\bibitem[Rutledge et al.(2002)]{Rut02} Rutledge, R.~E., Bildsten, L., Brown, E.~F., Pavlov, G.~G., Zavlin, V.~E.~2002, \apj, 577, 346

\bibitem[Servillat et al.(2008)]{Serv08} Servillat, M., Webb, N.~A., Barret, D. 2008, A\&A, 480, 397
  
\bibitem[Servillat et al.(2012)]{Serv12} Servillat, M., Heinke, C.~O., Ho, W.~C.~G., et al.~2012, MNRAS, 423, 1556  

\bibitem[Steiner et al.(2010)]{Stein10} Steiner, A.~W., Lattimer, J.~M., \& Brown, E.~F.~2010, \apj, 722, 33

\bibitem[Tsang et al.(2012)]{Tsang12} Tsang, M.~B., Stone, J.~R., Camera, F.~2012, \prc, 86, 015803
  
\bibitem[Walsh et al.(2015)]{Walsh15} Walsh, A.~R., Cackett, E.~M., Bernardini, F.~2015, \mnras, 449, 1238
  
\bibitem[Wang et al.(2013)]{Wang13} Wang, Z., Breton, R.~P., Heinke, C.~O., Deloye, C.~J., Zhong, J.~2013, \apj, 765, 151
  
\bibitem[Watkins et al.(2015)]{Wat15}
Watkins, L.~L., van der Marel, R.~P., Bellini, A., Anderson, J.~2015, \apj, in press (eprint arXiv:1509.00513)
  
\bibitem[Webb \& Barret(2007)]{Webb07} Webb, N.~A., \& Barret, D.~2007, \apj, 671, 727

\bibitem[Wijnands et al.(2013)]{Wij13} Wijnands, R., Degenaar, N., \& Page, D.~2013, MNRAS, 432, 2366
  
\bibitem[Wilms et al.(2000)]{Wilms00} Wilms, J., Allen, A., \& McCray, R.~2000, \apj, 542, 914

\bibitem[Woodley et al.(2012)]{Woodley12} Woodley, K.~A., Goldsbury, R., Kalirai, J. S., et al.~2012, AJ, 143, 50

\bibitem[Xu et al.(2014)]{Xu14} Xu, J., van Dyk, D.~A., Kashyap, V.~L., et al.~2014, \apj, 794, 97
  
\bibitem[Yakovlev \& Pethick(2004)]{Yak04} Yakovlev, D.~G., \& Pethick, C.~J. 2004, ARA\&A, 42, 169
  
\bibitem[Zavlin et al.(1996)]{Zavlin96} Zavlin, V.~E., Pavlov, G.~G., \& Shibanov, Yu.~A.~1996, A\&A, 315, 141

\bibitem[Zavlin \& Pavlov(2002)]{Zavlin02} Zavlin, V.~E., \& Pavlov,
  G.~G.~2002, in Neutron Stars, Pulsars, and Supernova Remnants,
  ed.~W.~Becker, H. Lesch, \& J. Tr\"umper (Garching bei M\"unchen:
  Max-Plank-Institut f\"ur extraterrestrische Physik), 263
  
\end{thebibliography}
\end{document}